\documentclass[onecollarge]{svjour2}       
\usepackage{graphicx}
\begin{document}

\title{Tunneling time and superposition principle}


\author{Nikolay L. Chuprikov 
}


\institute{N. L. Chuprikov \at
              Tomsk State Pedagogical University, 634041, Tomsk, Russia \\
              \email{chnl@tspu.edu.ru}           
}

\date{Received: date / Accepted: date}

\maketitle
\begin{abstract}

We show that scattering a quantum particle on a one-dimensional potential barrier as well as scattering the electromagnetic wave on a
quasi-one-dimensional layered structure (both represent scattering problems with one 'source' and two 'sinks') violate the superposition
principle; the role of nonlinear elements is played here by the potential barrier and the layered structure, splitting the incident (probability
and electromagnetic) wave into two parts (transmitted and reflected). This explains why all attempts to solve the tunneling time problem within
the framework of the standard (linear) models of these processes, both in quantum mechanics and in classical electrodynamics, have been
unsuccessful. We revise the traditional formulation of the superposition principle, present a new (nonlinear) wave model, by the example of the
quantum-mechanical scattering process, and show that concepts of the tunneling time developed on its basis are free from the Hartman paradox.

\end{abstract}

\newcommand{\Api}{A^{in}}
\newcommand{\Ami}{B^{in}}
\newcommand{\Apo}{A^{out}}
\newcommand{\Amo}{B^{out}}
\newcommand {\uta} {\tau_{tr}}
\newcommand {\utb} {\tau_{ref}}
\newcommand{\ppp}{\mbox{\hspace{5mm}}}
\newcommand{\ooo}{\mbox{\hspace{3mm}}}
\newcommand{\ooa}{\mbox{\hspace{1mm}}}
\newcommand{\ppa}{\mbox{\hspace{25mm}}}
\newcommand{\ppb}{\mbox{\hspace{35mm}}}
\newcommand{\ppc}{\mbox{\hspace{10mm}}}

\section{Introduction} \label{intro}

Scattering a quantum particle on a one-dimensional potential barrier, where the particle has two mutually exclusive possibilities -- either to
pass (tunnel) through the barrier or to be reflected from it -- is one of the simplest scattering problems in quantum mechanics (QM). But the
simplicity of this "two-channel" scattering process is deceptive, since the study of its temporal aspects leads to the tunneling time problem
(TTP), with its key question "How long does it take to tunnel through the barrier?", which remains unresolved up to date. Experts on this problem
(see, for example, the reviews \cite{Ha2,La1,Olk1,Nus,Mug,Ste,Win}) analyze a huge number of candidates for the role of the tunneling time and
show that none of them is really suitable for this purpose because, as was said in \cite{Ha2}, "All [the known tunneling-time concepts] have been
found to suffer one logical flaw or another, flaws sufficiently serious that must be rejected".

Now we can also add that all the hypotheses expressed in \cite{Ha2,La1,Olk1,Nus,Mug,Ste,Win} regarding the underlying cause that makes this
problem intractable also turned out to be far from the truth (this is indirectly explained by the fact that all the recent tunneling-time
approaches (see, for example, \cite{Lun,SoAh,Gros,Xue,Zhi}), like the previous ones, "suffer one logical flaw or another" (see, e.g.,
\cite{Soko})). And, perhaps, only those experts who linked the difficulties of solving the TPP with the fundamental problems of the QM itself were
the closest thing to the truth. For example, as was stressed in \cite{Mug}, "A very important aspect, not technical but fundamental, is that the
existing solutions [of the TTP], or even the identification of the difficulties, are closely linked to particular interpretations of quantum
mechanics\dots [Thus,] no simple, unambiguous, and quick resolution of all deep questions involved may be expected, since these concern our
understanding of the emergence of the classical world of events from the quantum world of possibilities."

But, nevertheless, even this point of view contains only a part of the truth. It does not take into account the fact that the TTP appears not only
in QM, but also in classical electrodynamics (CED). The same paradoxical Hartman effect and negative tunneling times appear for the monochromatic
electromagnetic wave scattering on a quasi-one-dimensional layered structure. Thus, the way to resolve the TTP and associated paradoxes should be
common in QM and CED. That is, "our [not] understanding of the emergence of the classical world of events from the quantum world of possibilities"
is not those root cause that hinders the solution of the TTP. The TTP is not only a problem of QM. All the difficulties arising in solving this
problem result, rather, from our (not) understanding of the role of the superposition principle in the scattering processes, in which the
(probability or electromagnetic) wave emitted by a single 'source' is then divided by a scatterer (the potential barrier or the layered structure)
onto two portions which move toward 'sinks' located on different side of the scatterer.

Our research shows that this reason lies in the standard quantum-mechanical and electrodynamic models of these two scattering processes. So far,
despite the existing disagreements between researchers dealing with the TTP, no one questions these models. In particular, no one questions the
applicability of the superposition principle to these processes, both in QM and CED. At the same time, as will be shown below, a one-dimensional
potential barrier and a layered structure serve for a quantum particle and the electromagnetic wave, respectively, as nonlinear elements that make
the superposition principle inapplicable to these scattering processes. We shall show this by the example of the quantum-mechanical scattering
problem, focusing only on the wave aspects of this process.

\section{Tunneling and superposition principle} \label{superpos}

We begin our analysis with the stationary Schr\"{o}dinger equation that describes the process of scattering a particle on a one-dimensional
potential barrier $V(x)$, nonzero in the spatial interval $[a,b]$. For the particle with the energy $E=\hbar^2k^2/2m$, where $m$ is the particle's
mass and $\hbar k$ is its momentum, the general solution $\psi(x,k)$ outside the interval $[a,b]$ can be written in the form
\begin{eqnarray} \label{1s}
\psi(x,k)=\left\{
\begin{array}{rl}
A_l e^{ikx}+B_le^{-ikx}:\ppp x\leq a\\
A_r e^{ikx}+ B_r e^{-ikx}:\ppp x\geq b
\end{array} \right.
\end{eqnarray}
$k=\sqrt{2mE}/\hbar$. The main requirement imposed on a searched-for wave function $\psi(x,k)$ is that this function and its first $x$-derivative
must be continuous everywhere on the $OX$-axis. Thus, this function must obey four real {\it linear} conditions of continuity (two real conditions
for the complex-valued $\psi(x,k)$ and two real conditions for its first $x$-derivative) at the points $x=a$ and $x=b$, where $V(x)$ is
discontinuous.

According to the transfer matrix approach, the wave amplitudes in Exps. (\ref{1s}) are linked by the transfer matrix $\textbf{Y}$:
\begin{eqnarray} \label{2s}
\left(
\begin{array}{rl}
A_l\\ B_l
\end{array} \right) =\textbf{Y}
\left(
\begin{array}{rl}
A_r \\ B_r
\end{array} \right);\ppp
\textbf{Y}=\left(
\begin{array}{rl}
q & p \\ p^* & q^*
\end{array} \right);\ppp |q|^2-|p|^2=1;
\end{eqnarray}
the matrix elements $q$ and $p$ are uniquely determined by the potential function $V(x)$. Besides, two (independent) amplitudes are determined
here by the boundary conditions at the regions $x<a$ and $x>b$. For example, for a particle impinging on the barrier from the left, the standard
boundary conditions are as follows: $A_l=1$, $B_r=0$. In this case, from Eq. (\ref{2s}) it follows that
\begin{eqnarray} \label{3s}
\psi(x,k)=\left\{
\begin{array}{rl}
e^{ikx}+\frac{p^*}{q}e^{-ikx}:\ppp x\leq a\\
\frac{1}{q} e^{ikx}:\ppp x\geq b
\end{array} \right.
\end{eqnarray}

Thus, the in- and out-asymptotes of the corresponding time-dependent solution $\psi(x,t)$ are
\begin{eqnarray} \label{4s}
\psi_{inc}(x,t)=\frac{1}{\sqrt{2\pi}}\int_{-\infty}^\infty {\cal{A}}(k)e^{-iE(k)t/\hbar}dk;\ppp \psi_{out}(x,t)=\psi_{tr}(x,t)+\psi_{ref}(x,t);\\
\psi_{tr}(x,t)=\frac{1}{\sqrt{2\pi}}\int_{-\infty}^\infty \frac{{\cal{A}}(k)}{q(k)}e^{i[kx-E(k)t/\hbar]}dk,\ooo
\psi_{ref}(x,t)=\frac{1}{\sqrt{2\pi}}\int_{-\infty}^\infty {\cal{A}}(k)\frac{p^*(k)}{q(k)}e^{-i[kx+E(k)t/\hbar]}dk;\nonumber
\end{eqnarray}
where ${\cal{A}}(k)$ is a complex-valued function determined by the initial condition (it is assumed that for 'physical initial states' the
function ${\cal{A}}(k)$ belongs to the Schwartz space). That is, in the limit $t\to -\infty$ the time-dependent solution $\psi(x,t)$ approaches
the left in-asymptote $\psi_{inc}(x,t)$ that represents a single incident wave packet, while in the limit $t\to +\infty$ the wave function
$\psi(x,t)$ approaches the out-asymptote $\psi_{out}(x,t)$ that represents the superposition of the right out-asymptote $\psi_{tr}(x,t)$ (a
transmitted wave packet) and the left out-asymptote $\psi_{ref}(x,t)$ (a reflected wave packet).

It is generally accepted that this scattering process a priori respects the superposition principle, and this standard (linear) quantum-mechanical
model is internally consistent. But is it?

As is seen, the main peculiarity of this scattering problem is that it involves one 'source' and two 'sinks'. With taking into account that in QM
there is no internal difference between 'source' and 'sink' (because the Schr\"{o}dinger's dynamics is reversible in time), this problem can also
be considered as a scattering problem with one 'sink' and two 'sources'. That is, if this process were to respect the superposition principle,
then the fact that the out-asymptote $\psi_{out}(x,t)$ is the superposition of the wave packets $\psi_{tr}(x,t)$ and $\psi_{ref}(x,t)$ would mean
that the asymptote $\psi_{inc}(x,t)$ represents the superposition of their 'predecessors'; accordingly, the incident wave (see Exp. (\ref{3s}))
represents a superposition of two incident waves, one of which is connected by cause-effect relations to the transmitted wave, and the other is
associated with the reflected wave.

But the standard model, with its linear continuity conditions, does not allow for a semitransparent potential barrier the existence of stationary
solutions which would have one incoming wave and one outgoing wave. That is, it does not imply the existence of incident waves for each
subprocess. In the final analysis, it does not imply the individual description of the transmission (tunneling) and reflection subprocesses at all
stages of scattering.

At first glance, we may fill this gap and resolve this problem, remaining within the standard model, with making use of two stationary solutions
$\psi_{(1)}(x,k)$ and $\psi_{(2)}(x,k)$; the first has a single outgoing wave being the transmitted wave $\psi_{tr}(x,k)$, and the second has a
single outgoing wave being the reflected wave $\psi_{ref}(x,k)$:
\begin{eqnarray} \label{5s}
\psi_{(1)}(x,k)=\left\{
\begin{array}{rl}
\frac{1}{|q|^2} e^{ikx}:\ppp x\leq a\\
\frac{1}{q} e^{ikx}-\frac{p^*}{|q|^2} e^{-ikx}:\ppp x\geq b
\end{array} \right.;\ppp \psi_{(2)}(x,k)=\left\{
\begin{array}{rl}
|\frac{p}{q}|^2 e^{ikx}+ \frac{p^*}{q} e^{-ikx}:\ppp x\leq a\\
\frac{p^*}{|q|^2} e^{-ikx}:\ppp x\geq b
\end{array} \right.
\end{eqnarray}
As is seen, they are such that $\psi_{(1)}(x,k)+\psi_{(2)}(x,k)=\psi(x,k)$. In this case the superposition of the incident waves $\psi_{(1)}(x,k)$
and $\psi_{(2)}(x,k)$ in the spatial domain $x\geq b$ is destructive, resulting in their complete disappearance. While the superposition of their
incident waves in the region $x\leq a$ is constructive, giving the incident wave of the initial solution $\psi(x,k)$ (see (\ref{3s})).

But all this does not at all mean that the incident wave $\frac{1}{|q|^2} e^{ikx}$ of $\psi_{(1)}(x,k)$ is causally connected to the transmitted
wave $\frac{1}{q} e^{ikx}$, and the incident wave $|\frac{p}{q}|^2 e^{ikx}$ of $\psi_{(2)}(x,k)$ is causally connected to the reflected wave
$\frac{p^*}{q} e^{-ikx}$. This is so, because the probability current densities corresponding to the incoming and outgoing waves, in each pair of
wave functions, differ from each other.

This fact as well as the fact that the superposition of the wave functions $\psi_{(1)}(x,k)$ and $\psi_{(2)}(x,k)$ (each of them is associated
with two 'sources' and one 'sink') leads to their cardinal reconstruction (their superposition -- the wave function $\psi(x,k)$ -- is associated
with one (left) 'source' and two (left and right) 'sinks') mean that this scattering process violates the superposition principle and its standard
(linear) model is internally inconsistent. Thus, the standard formulation of the superposition principle as well as the standard ({\it linear})
model of this quantum mechanical scattering process must be revised.

Of course, revising is not needed in the particular case, when the potential barrier is either fully transparent or fully opaque (a one-channel
scattering). We must imply that, as before, a coherent superposition of two or more wave functions that describe possible states of a pure quantum
ensemble moving in the physical context formed by the left 'source' and (right or left) 'sink' gives a new state of this pure ensemble; in this
case, the interference between states forming this superposition -- a new possible state determined by this physical context -- makes them
indistinguishable; characteristic times and other physical quantities can be defined only for this new state of the pure ensemble.

However, in the general case (when both left and right "sinks" are involved), a new formulation of the superposition principle should imply that
the coherent superposition of pure states associated with different (left and right) "sinks" is a state of a mixture of two pure quantum ensembles
determined by different physical contexts: two pure states forming this superposition remain distinguishable, despite the interference between
them; characteristic times and other physical quantities can be now defined only for these two pure states, but not for their mixture. We have to
stress that this result is in a full agreement with our recent study \cite{Ch7}, where we have presented a superselection rule which restricts the
validity of the superposition principle in the rigged Hilbert space of states of a particle scattering on a one-dimensional potential barrier. All
this requires a new (nonlinear) model of this two-channel scattering process which would describe the individual dynamics of the transmitted and
reflected wave packets at all stages of scattering.

So, the main results of this section, extended onto the corresponding scattering process in CED, can be summarized as follows:
\begin{itemize}
\item[$\bullet$] "\dots our [non] understanding of the emergence of the classical world of events from the quantum world of possibilities" is not the
root cause that makes the solution of the TTP impossible; rather, our false understanding of the role of the principle of superposition in the
problem of scattering a classical electromagnetic wave on a quasi-one-dimensional layered structure makes it impossible both the solution of the
TTP and "\dots our understanding of the emergence of the classical world of events from the quantum world of possibilities";
\item[$\bullet$] scattering a quantum particle on a one-dimensional potential barrier
and scattering the electromagnetic wave on a quasi-one-dimensional layered structure are {\it nonlinear} phenomena;
\item[$\bullet$] in both cases the role of nonlinear elements is played by a potential barrier and layered structure that split the incident wave
(wave packet) into the transmitted and reflected waves (wave packets);
\item[$\bullet$] the standard models of these two scattering processes, both in QM and in CED, are linear and, thus, they do not give an adequate
description of these processes; the adequate (nonlinear) model of each of these two two-channel scattering processes must answer the question of
how to uniquely represent the incident wave associated with the whole process in the form of a superposition of the incident wave, which would be
causally related only to the transmitted wave, and the incident wave, which would be causally related only to the reflected wave;
\item[$\bullet$] for each subprocess (transmission and reflection), a causal relationship between the incident wave and the corresponding
outgoing wave can be realized only on the basis of {\it nonlinear} continuity conditions that have yet to be formulated.
\end{itemize}

Note that the relevant nonlinear tunneling models, both in QM and CED, have already been developed, based on intuitive considerations, and
presented in our articles \cite{Ch1,Ch2} and \cite{Ch9}, respectively. These models give the 'subprocess' wave functions (SWFs) which allow one to
describe the transmission and reflection subprocesses at all stages of scattering. And on their basis the transmission (tunneling) and reflection
times have been defined. In this paper, a new (nonlinear) quantum-mechanical model is presented on a more rigorous basis with the addition of new
important details.

\section{Standard model of scattering a particle on a system of two identical rectangular potential barriers} \label{back}

In connection with the subsequent analysis of the generalized Hartman paradox, a standard quantum mechanical description of tunneling will be
presented by the example of scattering a quantum particle on a one-dimensional system of two identical rectangular potential barriers (when the
gap between barriers is zero, this model describes the tunneling of a particle through a single rectangular potential barrier). In doing so, we
will use our version \cite{Ch8} of the transfer matrix approach.

Let a particle with a given momentum $\hbar k$ ($k>0$) impinge from the left on the system of two identical rectangular potential barriers located
at the intervals $[a_1,b_1]$ and $[a_2,b_2]$; $0<a_1<b_1<a_2<b_2$. The height of both barriers is $V_0$, $b_1-a_1=b_2-a_2=d$ is their width;
$L=a_2-b_1$ is the distance between the barriers; $b_2-a_1=D$ is the width of this two-barrier system.

The wave function $\Psi_{tot}(x,k)$ that describes the state of the quantum ensemble of such particles can be written as follows:
\begin{eqnarray} \label{1}
\Psi_{tot}(x,k)= \left\{
\begin{array}{rl}
e^{ikx}+B_{out}e^{ik(2a_1-x)}:\ppp x\in (-\infty, a_1]\\
A_{tot}^{(1)}\sinh[\kappa(x-a_1)]+B_{tot}^{(1)}\cosh[\kappa(x-a_1)]:\ppp x\in[a_1,b_1]\\
A_{tot}^{gap}\sin[k(x-x_c)]+B_{tot}^{gap}\cos[k(x-x_c)]:\ppp x\in[b_1,a_2]\\
A_{tot}^{(2)}\sinh[\kappa(x-b_2)]+B_{tot}^{(2)}\cosh[\kappa(x-b_2)]:\ppp x\in[a_2,b_2]\\
A_{out}e^{ik(x-D)}:\ppp x\in[b_2,\infty)
\end{array} \right.
\end{eqnarray}
here $\kappa=\sqrt{2m(V_0-E)}/\hbar$; $E=\hbar^2 k^2/2m$; $x_c=(b_2+a_1)/2$. We have to stress that the formalism presented here is valid not only
for $E<V_0$ (when the Hartman paradox appears in the opaque barrier limit $d\to\infty$) but also for $E\geq V_0$ (in this case, $\kappa$ is a
purely imaginary quantity).

According to \cite{Ch8}, for any semitransparent potential barrier located in the interval $[a,b]$, the elements $q$ and $p$ of the transfer
matrix $\textbf{Y}$ (see (\ref{2s})) can be presented in the form
\begin{eqnarray} \label{2}
q=\frac{1}{\sqrt{T_{(a,b)}}}\exp\left\{i\left[k(b-a)-J_{(a,b)}\right]\right\},\ooa p=i\sqrt{\frac{R_{(a,b)}}{T_{(a,b)}}} \ooa
\exp\left\{i\left[F_{(a,b)}-k(b+a)\right]\right\},
\end{eqnarray}
where (see \cite{Ch8}) the transmission coefficient $T_{(a,b)}$ and phases $J_{(a,b)}$ and $F_{(a,b)}$ are determined either by explicit
analytical expressions (e.g., for the rectangular barrier and the $\delta$-potential) or by the recurrence relations (for many-barrier
structures); $R_{(a,b)}=1-T_{(a,b)}$; for any symmetric system of barriers, when $V(x-x_c)=V(x_c-x)$, the phase $F_{(a,b)}$ can take only two
values, either $0$ or $\pi$.

For the transfer matrices $\textbf{Y}_{two}$, $\textbf{Y}_1$ and $\textbf{Y}_2$ that describe this two-barrier system as well as its left and
right barriers, respectively, we have
\begin{eqnarray} \label{3}
\left(
\begin{array}{rl}
1\\ B_{out}e^{2ika_1}
\end{array} \right) =\textbf{Y}_{two}
\left(
\begin{array}{rl}
A_{out}e^{-ikD} \\ 0
\end{array} \right);\ppp \textbf{Y}_{two}=\textbf{Y}_1\textbf{Y}_2=\left(
\begin{array}{rl}
q_{two} & p_{two} \\ p^*_{two} & q^*_{two}
\end{array} \right),\ppp \textbf{Y}_n=\left(
\begin{array}{rl}
q_n & p_n \\ p^*_n & q^*_n
\end{array} \right)
\end{eqnarray}
where $q_n=q\cdot\exp[ik(b_n-a_n)]$, $p_n=ip\cdot\exp[-ik(b_n+a_n)]$ $(n=1,2)$;
\begin{eqnarray*}
q=\frac{e^{-iJ}}{\sqrt{T}},\ooo p=\sqrt{\frac{R}{T}}\ooa e^{iF};\ppp q_{two}=\frac{1}{\sqrt{T_{two}}}e^{i[k(b_2-a_1)-J_{two}]},\ooo
p_{two}=i\sqrt{\frac{R_{two}}{T_{two}}}\ooa e^{i[F_{two}-k(b_2+a_1)]}
\end{eqnarray*}
For rectangular barriers the 'one-barrier' parameters $T$, $J$ and $F$ are (see also \cite{Ch8})
\begin{eqnarray} \label{300}
T=\left[1+\theta_{(+)}^2 \sinh^2(\kappa d)\right]^{-1},\ooa J=\arctan\left(\theta_{(-)}\tanh(\kappa d)\right)+J^{(0)},\ooa
\theta_{(\pm)}=\frac{1}{2}\left(\frac{k}{\kappa}\pm \frac{\kappa}{k}\right);
\end{eqnarray}
$J^{(0)}=0$, if $\cosh(\kappa d)>0$; otherwise, $J^{(0)}=\pi$ (this can occur for $E\geq V_0$); $F=0$, if $\theta_{(+)}\sinh(\kappa d)>0$;
otherwise, $F=\pi$. From the latter it follows that the parameter $p$ is real. It can be rewritten in the form $p=\eta \sqrt{R/T}$; where
$\eta=+1$, if $\theta_{(+)}\sinh(\kappa d)>0$; otherwise, $\eta=-1$.

The 'two-barrier' parameters $T_{two}$, $J_{two}$ and $F_{two}$ are determined by Eq. (\ref{3}) (see also the recurrence relations for the
scattering parameters in \cite{Ch8}):
\begin{eqnarray} \label{4}
T_{two}^{-1}=1+4\frac{R}{T^2}\cos^2\chi,\ooo J_{two}=J+\arctan\left(\frac{1-R}{1+R}\tan\chi\right)+F_{two}^{(0)},\ooo F_{two}=F+F_{two}^{(0)};
\end{eqnarray}
here $\chi=J+kL$; $F_{two}^{(0)}=0$, if $\cos\chi\geq 0$; otherwise, $F_{two}^{(0)}=\pi$ (the piecewise constant function $F_{two}(k)$ is
discontinuous at the resonance points where $T_{two}=1$).

Now the coefficients in Exps. (\ref{1}) can be written in terms of these one-barrier and two-barrier parameters scattering. For this purpose it is
suitable to rewrite the wave function $\Psi_{tot}(x,k)$ in the interval $[b_1,a_2]$ in the form
$\Psi_{tot}(x,k)=\tilde{A}_{tot}^{gap}\exp(ikx)+\tilde{B}_{tot}^{gap}\exp(-ikx)$ where
\begin{eqnarray} \label{5}
\left(
\begin{array}{rl}
\tilde{A}_{tot}^{gap}\\ \tilde{B}_{tot}^{gap}
\end{array} \right) =\textbf{Y}_2
\left(
\begin{array}{rl}
A_{out}e^{-ikD)} \\ 0
\end{array} \right) =\textbf{Y}_1^{-1}
\left(
\begin{array}{rl}
1 \\ B_{out}e^{2ika_1}
\end{array} \right).
\end{eqnarray}
Since $A_{tot}^{gap}=i\left(\tilde{A}_{tot}^{gap}e^{ikx_c}-\tilde{B}_{tot}^{gap}e^{-ikx_c}\right)$ and
$B_{tot}^{gap}=\tilde{A}_{tot}^{gap}e^{ikx_c}+\tilde{B}_{tot}^{gap}e^{-ikx_c}$, from the first equality in (\ref{5}) it follows that
$A_{tot}^{gap}$ and $B_{tot}^{gap}$ in (\ref{1}) are determined by the expressions
\begin{eqnarray} \label{6}
A_{tot}^{gap}= -A_{out}P^*e^{ika_1},\ppp B_{tot}^{gap}=A_{out}Q^*e^{ika_1};
\end{eqnarray}
here $Q=q^*\exp(ikL/2)+ip\ooa \exp(-ikL/2)$, $P=iq^*\exp(ikL/2)+p\ooa\exp(-ikL/2)$. Besides, by 'sewing' the solutions (\ref{1}) at the points
$x=a_1$ and $x=b_2$, we obtain
\begin{eqnarray} \label{7}
A_{tot}^{(1)}=i (1-B_{out})\frac{k}{\kappa}\ooa e^{ika_1},\ooo B_{tot}^{(1)}=(1+B_{out})\ooa e^{ika_1};\ooo A_{tot}^{(2)}=i
A_{out}\frac{k}{\kappa}\ooa e^{ika_1},\ooo B_{tot}^{(2)}=A_{out}e^{ika_1}.\nonumber
\end{eqnarray}

The amplitudes $A_{out}$ and $B_{out}$ can be expressed either through the one-barrier parameters, with help of the second equality in (\ref{5}),
or through the two-barrier ones, with help of the relationship
\begin{eqnarray*}
\left(
\begin{array}{rl}
1 \\ B_{out}e^{2ika_1}
\end{array} \right)=\textbf{Y}_{two}
\left(
\begin{array}{rl}
A_{out}e^{-ikD} \\ 0
\end{array} \right).
\end{eqnarray*}
As a result, we have two equivalent forms for each amplitude,
\begin{eqnarray} \label{8}
A_{out}=\frac{1}{2}\left(\frac{Q}{Q^*}-\frac{P}{P^*}\right)=\sqrt{T_{two}}\ooa e^{iJ_{two}},\nonumber\\
B_{out}=-\frac{1}{2}\left(\frac{Q}{Q^*}+\frac{P}{P^*}\right)=-i\sqrt{R_{two}}\ooa e^{i(J_{two}-F_{two})}
\end{eqnarray}
(both the forms will be useful for developing a new model of this process.

\section{A new, nonlinear model of scattering a particle on a system of two identical rectangular potential barriers} \label{separ}

\subsection{Stationary wave functions for transmission and reflection} \label{tref}

According to \cite{Ch1}, for any symmetric two-barrier system the total wave function $\Psi_{tot}(x,k)$ to describe the whole scattering process
can be uniquely presented, for any values of $x$ and $k$, as a superposition of two SWFs $\psi_{tr}(x,k)$ and $\psi_{ref}(x,k)$ to describe the
transmission and reflection subprocesses, respectively. Both possess the following properties:
\begin{itemize}
\item[(a)]
$\psi_{tr}(x,k)+\psi_{ref}(x,k)=\Psi_{tot}(x,k)$;

\item[(b)] each SWF has only one outgoing wave and only one incoming wave; in this case the transmitted wave in (\ref{1}) serves as
the outgoing wave in $\psi_{tr}(x,k)$, while the reflected one represents the outgoing wave in $\psi_{ref}(x,k)$;

\item[(c)] the incoming wave and the corresponding outgoing wave of each SWF, extended into the barrier region, join together at some point
$x_{join}(k)$, where the SWF and the corresponding density of probability density are continuous.
\end{itemize}

From (c) it follows that the continuity conditions used to sew incoming (falling) and outgoing waves for each subprocess represent three (real)
continuity conditions: two (real) conditions for the complex-valued SWF itself and one (real) continuity condition for the corresponding
probability current density. Thus, these three (real) continuity conditions are {\it nonlinear} as it should be, according to a new formulation of
the superposition principle! In this case the first $x$-derivatives of both SWFs are discontinuous at the point $x_c$.

A simple analysis shows that for any {\it symmetric} two-barrier system $x_{join}(k)$ coincides, for any value of $k$, with the midpoint $x_c$ of
the barrier region: $x_c=(b_2+a_1)/2$. In this case, $\psi_{ref}(x,k)\equiv 0$ and $\psi_{tr}(x,k)\equiv \Psi_{tot}(x,k)$ for $x\geq x_c$. This
means that particles reflected by the symmetric two-barrier system do not enter into the region $x>x_c$, and the SWF $\psi_{ref}(x,k)$ is a
currentless wave function.

Calculations yield that in the region $x<x_c$ the wave function $\psi_{ref}(x,k)$ can be written as follows,
\begin{eqnarray} \label{102}
\psi_{ref}(x,k)= \left\{
\begin{array}{rl}
A^{in}_{ref}e^{ikx}+b_{out}e^{ik(2a_1-x)}:\ppp x\in (-\infty, a_1]\\
a_{ref}^{(1)}\sinh[\kappa(x-b_1)]+b_{ref}^{(1)}\cosh[\kappa(x-b_1)]:\ppp x\in[a_1,b_1]\\
a_{ref}^{gap}\sin[k(x-x_c)]:\ppp x\in[b_1,x_c]
\end{array} \right.
\end{eqnarray}
Again, as in Section \ref{back}, in order to find the amplitudes in these expressions it is suitable to rewrite the function $\psi_{ref}(x,k)$ in
the interval $[b_1,x_c]$ in the form $\psi_{ref}(x,k)=A_{ref}^{gap}\exp(ikx)+B_{ref}^{gap}\exp(-ikx)$. The amplitudes in this expression are
linked as follows
\begin{eqnarray} \label{103}
\left(
\begin{array}{rl}
A_{ref}^{gap}\\ B_{ref}^{gap}
\end{array} \right) =\textbf{Y}_1^{-1}
\left(
\begin{array}{rl}
A^{in}_{ref} \\ b_{out}e^{2ika_1}
\end{array} \right).
\end{eqnarray}
Then, making use of the relationships
\begin{eqnarray} \label{104}
a_{ref}^{gap}=i\left(A_{ref}^{gap}\ooa e^{ikx_c}-B_{ref}^{gap}\ooa e^{-ikx_c}\right),\ooo A_{ref}^{gap}\ooa e^{ikx_c}+B_{ref}^{gap}\ooa
e^{-ikx_c}=0
\end{eqnarray}
we find the unknown amplitudes in Exps. (\ref{102}).

From the second equality in (\ref{104}) it follows that $A^{in}_{ref}=-b_{out}Q^*/Q$. Then, taking into account Exps. (\ref{7}), we obtain
\begin{eqnarray} \label{105}
A^{in}_{ref}=b_{out}(b_{out}^*-a_{out}^*)=\sqrt{R_{two}}\left(\sqrt{R_{two}}+i\eta_{two}\sqrt{T_{two}}\right)\equiv \sqrt{R_{two}}\exp(i\lambda)
\end{eqnarray}
where $\eta_{two}=+1$, if $F_{two}=0$; otherwise, $\eta_{two}=-1$. This means that the phases of the incident waves in $\Psi_{tot}(x,k)$ and
$\psi_{ref}(x,k)$ differ from each other by the amount $\lambda=\eta_{two}\cdot \arctan\sqrt{T_{two}(k)/R_{two}(k)}$.

Then, taking into account, in (\ref{104}), Exps. (\ref{103}) and (\ref{105}), we obtain
\begin{eqnarray*}
a_{ref}^{gap}=-2P b_{out} a^*_{out}e^{ika_1}.
\end{eqnarray*}
And lastly, making use of the continuity conditions at the point $x=b_1$, we obtain
\begin{eqnarray*}
a_{ref}^{(1)}=\frac{k}{\kappa}a_{ref}^{gap}\cos\left(\frac{kL}{2}\right),\ppp b_{ref}^{(1)}=-a_{ref}^{gap}\sin\left(\frac{kL}{2}\right).
\end{eqnarray*}

Since $\psi_{ref}(x,k)$ is now known, we have $\psi_{tr}(x,k)=\Psi_{tot}(x,k)-\psi_{ref}(x,k)$. In particular,
\begin{eqnarray} \label{106}
A^{in}_{tr}=1-A^{in}_{ref}=\sqrt{T_{two}}\left(\sqrt{T_{two}}-i\eta_{two}\sqrt{R_{two}}\right)\equiv
\sqrt{T_{two}}\exp\left[i\left(\lambda-\eta_{two}\frac{\pi}{2}\right)\right].
\end{eqnarray}
As is seen, not only $A^{in}_{tr}(k)+A^{in}_{ref}(k)=1$, but also $|A^{in}_{tr}(k)|^2+|A^{in}_{ref}(k)|^2=1$. Besides,
\begin{eqnarray} \label{108}
|\psi_{tr}(x_c-x,k)|=|\psi_{tr}(x-x_c,k)|.
\end{eqnarray}

\subsection{Time-dependent wave functions for transmission and reflection} \label{alt}

Let us now proceed to the time-dependent process described by the wave packet
\begin{eqnarray} \label{200}
\Psi_{tot}(x,t)=\frac{1}{\sqrt{2\pi}}\int_{-\infty}^\infty {\cal{A}}(k)\Psi_{tot}(x,k)e^{-iE(k)t/\hbar}dk.
\end{eqnarray}
At this point (see also (\ref{4s})) we assume ${\cal{A}}(k)$ to be the Gaussian function ${\cal{A}}(k)=(2l_0^2/\pi)^{1/4}
\exp\left[-l_0^2(k-\bar{k})^2\right]$. In this case
\begin{eqnarray} \label{201}
\bar{x}_{tot}(0)=0,\ppp \bar{p}_{tot}(0)=\hbar\bar{k},\ppp\overline{x^2}_{tot}(0)=l_0^2;
\end{eqnarray}
hereinafter, for any observable $F$ and time-dependent localized state $\Psi^A_B(t)$
\[\bar{F}^A_B(t)=\frac{<\Psi^A_B(t)|\hat{F}|\Psi^A_B(t)>}{<\Psi^A_B(t)|\Psi^A_B(t)>}\] (if $\bar{F}^A_B(t)$ is constant its argument will be omitted). We
assume that the parameters $l_0$ and $\bar{k}$ obey the conditions for the scattering process: the rate of scattering the transmitted and
reflected wave packets is assumed to exceed the rate of widening each packet; so that the transmitted and reflected wave packets non-overlap each
other. We also assume that the origin of coordinates, from which the "center of mass"\/ (CM) $\bar{x}_{tot}$ of the wave packet $\Psi_{tot}(x,t)$
starts, lies sufficiently far from the left boundary of the two-barrier system: $a_1\gg l_0$.

Besides, let the expression
\begin{eqnarray}  \label{202}
\psi_{tr,ref}(x,t)=\frac{1}{\sqrt{2\pi}}\int_{-\infty}^\infty {\cal{A}}(k)\psi_{tr,ref}(x,k)e^{-iE(k)t/\hbar}dk
\end{eqnarray}
give the wave functions $\psi_{tr}(x,t)$ and $\psi_{ref}(x,t)$ to describe, respectively, the time-dependent transmission and reflection
subprocesses. It is evident (see the requirement (a) in Section \ref{tref}) that the sum of these two functions yields, at any value of $t$, the
total wave function $\Psi_{tot}(x,t)$,
\begin{eqnarray} \label{203}
\Psi_{tot}(x,t)=\psi_{tr}(x,t)+\psi_{ref}(x,t).
\end{eqnarray}

So, at the first stage, the process is described by the incident packet
\begin{eqnarray*}
\Psi_{tot}(x,t)\simeq\Psi_{tot}^{inc}(x,t)= \frac{1}{\sqrt{2\pi}}\int_{-\infty}^\infty {\cal{A}}(k)\exp[i(kx-E(k)t/\hbar)]dk,
\end{eqnarray*}
and its transmission and reflection subprocesses are described by the wave packets
\begin{eqnarray*}
\psi_{tr,ref}\simeq\psi_{tr,ref}^{inc}= \frac{1}{\sqrt{2\pi}}\int_{-\infty}^\infty A^{in}_{tr,ref}(k){\cal{A}}(k)\exp[i(kx-E(k)t/\hbar)]dk.
\end{eqnarray*}
Considering Exps. (\ref{105}) and (\ref{106}) for the amplitudes of the incident waves in $\psi_{tr}(x,k)$ and $\psi_{ref}(x,k)$, it is easy to
show that
\begin{eqnarray} \label{204}
{\bar{x}}_{tr}^{inc}(0)=-{\overline{\lambda^\prime(k)}}_{tr}^{inc}\equiv -\frac{\int_{-\infty}^\infty \lambda^\prime(k) T_{two}(k)
|{\cal{A}}(k)|^2 dk}{\int_{-\infty}^\infty T_{two}(k) |{\cal{A}}(k)|^2 dk},\\
{\bar{x}}_{ref}^{inc}(0)=-{\overline{\lambda^\prime(k)}}_{ref}^{inc}\equiv -\frac{\int_{-\infty}^\infty
\lambda^\prime(k)R_{two}(k)|{\cal{A}}(k)|^2 dk}{\int_{-\infty}^\infty R_{two}(k)|{\cal{A}}(k)|^2 dk};\nonumber
\end{eqnarray}
the prime denotes the derivative on $k$. That is, in the general case the CMs of the wave packets $\Psi_{tot}(x,t)$, $\psi_{tr}(x,t)$ and
$\psi_{ref}(x,t)$ start at $t=0$ from the different spatial points!

Similarly, for the final stage of the process
\begin{eqnarray*}
\psi_{tr}\simeq\psi_{tr}^{out}= \frac{1}{\sqrt{2\pi}}\int_{-\infty}^\infty {\cal{A}}(k)a_{out}(k)e^{i[k(x-D)-E(k)t/\hbar]}dk,\\
\psi_{ref}\simeq\psi_{ref}^{out}= \frac{1}{\sqrt{2\pi}}\int_{-\infty}^\infty {\cal{A}}(k)b_{out}(k)e^{i[k(2a_1-x)-E(k)t/\hbar]}dk.
\end{eqnarray*}
Thus, since $|A^{in}_{tr}(k)|^2=|a_{out}(k)|^2=T_{two}(k)$ and $|A^{in}_{ref}(k)|^2=|b_{out}(k)|^2=R_{two}(k)$ (see (\ref{8}), (\ref{105}) and
(\ref{106})), for the initial and final stages of scattering we have
\begin{eqnarray*}
\langle\psi_{tr}^{inc}|\psi_{tr}^{inc}\rangle=\langle\psi_{tr}^{out}|\psi_{tr}^{out}\rangle=\int_{-\infty}^\infty T_{two}(k)|{\cal{A}}(k)|^2dk
\equiv {\textbf{T}}_{as},\\ \langle\psi_{ref}^{inc}|\psi_{ref}^{inc}\rangle=\langle\psi_{ref}^{out}|\psi_{ref}^{out}\rangle =\int_{-\infty}^\infty
R_{two}(k)|{\cal{A}}(k)|^2dk\equiv {\textbf{R}}_{as}.
\end{eqnarray*}

In its turn, since $T_{two}(k)+R_{two}(k)=1$ and $\langle\Psi_{tot}|\Psi_{tot}\rangle=\int_{-\infty}^\infty |{\cal{A}}(k)|^2dk=1$, from the above
it follows that the sum of the constant norms ${\textbf{T}}_{as}$ and ${\textbf{R}}_{as}$ is equal to unity:
\begin{eqnarray} \label{205}
{\textbf{T}}_{as}+{\textbf{R}}_{as}=1.
\end{eqnarray}
From the fact that the transmission and reflection subprocesses obey, at both these stages of scattering, the probabilistic "either-or" rule
(\ref{205}) it follows that they behave as {\it alternative} subprocesses at both the stages, despite interference between them at the first
stage. We can add to the equality (\ref{205}) that
\begin{eqnarray*}
\langle\psi_{tr}^{inc}|\psi_{ref}^{inc}\rangle=\int_{-\infty}^\infty |{\cal{A}}(k)|^2\left[A^{in}_{tr}(k)\right]^*A^{in}_{ref}(k)dk =
i\int_{-\infty}^\infty |{\cal{A}}(k)|^2 \eta_{two}(k)\sqrt{T_{two}(k)R_{two}(k)}dk
\end{eqnarray*}
(the piece-wise constant function $\eta_{two}(k)$ is defined in (\ref{105})). Thus,
$\langle\psi_{tr}^{inc}|\psi_{ref}^{inc}\rangle+\langle\psi_{ref}^{inc}|\psi_{tr}^{inc}\rangle=0$.

At the very stage of scattering, when the wave packet $\psi_{tr}(x,t)$ crosses the point $x_c$, the norm
${\textbf{T}}=\langle\psi_{tr}|\psi_{tr}\rangle$ can change. The point is that the nonlinear continuity conditions (a)-(c) in Section \ref{tref}
guarantee the balance between the input $I_{tr}(x_c-0,k)$ and output $I_{tr}(x_c+0,k)$ probability flows only for the {\it stationary} waves
$\psi_{tr}(x,k)$, of which the wave packet $\psi_{tr}(x,t)$ is built. However, for the packet itself the interference between the main 'harmonic'
$\psi_{tr}(x,\bar{k})$ and 'subharmonics' $\psi_{tr}(x,k)$, whose first derivatives are discontinuous at the point $x_c$, leads to the imbalance
between the input and output flows at the point $x_c$: $d\textbf{T}/dt=I_{tr}(x_c+0,t)-I_{tr}(x_c-0,t)\neq 0$. Since the role of subharmonics is
essential only at the leading and trailing fronts of the wave-packet, this effect takes place only when these fronts cross the midpoint $x_c$.
Otherwise, ${\textbf{T}}+{\textbf{R}}\approx 1$ even at the very stage of scattering, when the CM of the wave packet $\psi_{tr}(x,t)$ moves inside
the barrier region.

Note that the total alteration of the norm $\textbf{T}$, accumulated in the course of the scattering process, is zero. As regards $\textbf{R}$,
this norm remains constant even at the very stage of scattering: $\textbf{R}\equiv\textbf{R}_{as}$. This follows from the fact that
$I_{ref}(x_c+0,t)=I_{ref}(x_c-0,t)=0$ since $\psi_{ref}(x_c,t)=0$ for any value of $t$.

\section{Local and asymptotic group ("phase") times for transmission and reflection} \label{as}

Now, when the dynamics of the sub-processes of transmission and reflection at all stages of scattering became known, we can proceed to study
temporal aspects of each subprocess. We begin with the presentation of the local (exact) and asymptotic (extrapolated) group times for
transmission and reflection. For example, the local transmission group time $\tau_{tr}^{loc}$ that characterizes the dynamics of the CM of the
wave packet $\psi_{tr}(x,t)$ inside the region $[a_1,b_2]$ is defined as follows (see \cite{Ch1}): $\tau_{tr}^{loc}=t_{tr}^{exit}-t_{tr}^{entry}$,
where $t_{tr}^{entry}$ and $t_{tr}^{exit}$ are such instants of time that
\begin{eqnarray*}
\bar{x}_{tr}(t_{tr}^{entry}) =a_1,\ppp \bar{x}_{tr}(t_{tr}^{exit})=b_2.
\end{eqnarray*}
Similarly, for reflection $\tau_{ref}^{loc}=t_{ref}^{exit}-t_{ref}^{entry}$, where $t_{ref}^{entry}$ and $t_{ref}^{exit}$ are two different roots,
if any, of the same equation ($t_{ref}^{entry}<t_{ref}^{exit}$):
\begin{eqnarray*}
\bar{x}_{ref}(t_{ref}^{entry}) =a_1,\ppp \bar{x}_{ref}(t_{ref}^{exit})=a_1.
\end{eqnarray*}
If this equation has no more than one root, $\tau_{ref}^{loc}=0$.

Note that the local group times $\tau_{tr}^{loc}$ and $\tau_{ref}^{loc}$ do not give a complete description of the temporal aspects of each
subprocess, because the two-barrier system affects the subensembles of transmitted and reflected particles not only when the CMs of the wave
packets $\psi_{tr}(x,t)$ and $\psi_{ref}(x,t)$ move in the region $[a_1,b_2]$. Of importance is also to define the {\it asymptotic} group times to
describe these subprocesses in the asymptotically large spatial region $[0,b_2+\Delta X]$ where $\Delta X\gg l_0$ .

In doing so, we have to take into account that each wave packet does not interact with the system when its CM is at the boundaries of this spatial
interval. That is, the {\it asymptotic} transmission time can be defined in terms of the transmitted $\psi_{tr}^{out}$ and to-be-transmitted
$\psi_{tr}^{inc}$ wave packets. Similarly, the asymptotic reflection time can be introduced in terms of the wave packets $\psi_{ref}^{out}$ and
$\psi_{ref}^{inc}$.

We begin with the transmission subprocess. For the CM's position $\bar{x}_{tr}(t)$ at the initial stage of scattering we have (see also
(\ref{204}))
\begin{eqnarray}  \label{301}
\bar{x}_{tr}(t)\simeq{\bar{x}}_{tr}^{inc}(t)=\frac{\hbar {\bar{k}}_{tr}}{m}t-{\overline{\lambda^\prime(k)}}_{tr}^{inc};
\end{eqnarray}
here ${\bar{k}}_{tr}={\bar{k}}_{tr}^{out}={\bar{k}}_{tr}^{inc}$. At the final stage
\begin{eqnarray*}
\bar{x}_{tr}(t)\simeq{\bar{x}}_{tr}^{out}(t)=\frac{\hbar {\bar{k}}_{tr}}{m}t-{\overline{J_{two}^\prime(k)}}_{tr}^{out} +D.
\end{eqnarray*}
Thus, the time $\tau^{gr}_{tr}(0,b_2+\Delta X)$ spent by the CM of $\psi_{tr}(x,t)$ in $[0,b_2+\Delta X]$ is
\begin{eqnarray*}
\tau^{gr}_{tr}(0,b_2+\Delta X)\equiv t_{arr}^{tr}-t_{dep}^{tr}=\frac{m}{\hbar \bar{k}_{tr}}\left[{\overline{J_{two}^\prime(k)}}_{tr}^{out}
-{\overline{\lambda^\prime(k)}}_{tr}^{inc}+a_1+\Delta X \right],
\end{eqnarray*}
where the arrival time $t_{arr}^{tr}$ and the departure time $t_{dep}^{tr}$ obey the equations
\[{\bar{x}}_{tr}^{inc}(t_{dep}^{tr})=0;\ooo {\bar{x}}_{tr}^{out}(t_{arr}^{tr})=b_2+\Delta X.
\]
The quantity $\tau_{tr}^{as}=\tau^{gr}_{tr}(a_1,b_2)$, which is associated with the region $[a_1,b_2]$, will be referred to as the asymptotic
(extrapolated) transmission group time:
\begin{eqnarray} \label{302}
\tau^{as}_{tr}=\frac{m}{\hbar \bar{k}_{tr}}\left[{\overline{J_{two}^\prime(k)}}_{tr}^{out} -{\overline{\lambda^\prime(k)}}_{tr}^{inc}\right].
\end{eqnarray}
Similarly, for reflection we have
\begin{eqnarray} \label{303}
\tau^{as}_{ref}=\frac{m}{\hbar \bar{k}_{ref}}\left[{\overline{J_{two}^\prime(k)}}_{ref}^{out} -{\overline{\lambda^\prime(k)}}_{ref}^{inc}\right];
\end{eqnarray}
${\bar{k}}_{ref}^{inc}=-{\bar{k}}_{ref}^{out}={\bar{k}}_{ref}$.

Let us now consider narrow (in $k$-space) wave packets (in this case we will omit the upper line in the notation $\bar{k}$):
\begin{eqnarray} \label{3029}
a_1, \Delta X\gg l_0\gg D.
\end{eqnarray}
Now Exps. (\ref{302}) and (\ref{303}) give the same value:
\begin{eqnarray} \label{3030}
\tau^{as}_{tr}(k)=\tau^{as}_{ref}(k)\equiv\tau_{as}(k)=\frac{m}{\hbar k}\left[J_{two}^\prime(k)-\lambda^\prime(k)\right];\\
t^{tr}_{dep}(k)=t^{ref}_{dep}(k)\equiv t_{dep}(k)=m\lambda^\prime(k)/\hbar k; \nonumber\\ {\bar{x}}_{tr}^{inc}(0)={\bar{x}}_{ref}^{inc}(0)\equiv
x_{start}=-\lambda^\prime(k).\nonumber
\end{eqnarray}
Here (see Exps. (\ref{4}) and (\ref{105}))
\begin{eqnarray*}
J_{two}^{\prime}=J^\prime+\frac{T_{two}}{T^2}\left[T(1+R)\left(J^\prime+L\right)+
T^\prime\sin[2(J+kL)]\right],\\
\lambda^{\prime}=2\eta \frac{T_{two}}{\sqrt{R}\ooa T^2}\left[T^\prime(1+R)\cos(J+kL)+2RT(J^\prime+L)\sin(J+kL)\right].
\end{eqnarray*}
These expressions are valid for any symmetric two-barrier system. For rectangular barriers we can obtain explicit expressions for one-barrier
functions. Namely, from Exps. (\ref{300}) it follows that
\begin{eqnarray*}
J^\prime=\frac{T}{\kappa}\left[\theta^2_{(+)}\sinh(2\kappa d)+\theta_{(-)}\kappa d\right],\ooo
T^\prime=2\theta^2_{(+)}\frac{T^2}{\kappa}\left[2\theta_{(-)}\sinh^2(\kappa d) +\kappa d \sinh(2\kappa d)\right].
\end{eqnarray*}

Note that the corresponding expressions for the Wigner phase time $\tau_{ph}$ \cite{Wig} has been obtained in \cite{Ol3}. In our notations it can
be written as $\tau_{ph}=mJ_{two}^\prime(k)/\hbar k$. This concept is based on the assumption that transmitted particles start, on average, from
the point $\bar{x}_{tot}(0)$ (\ref{201}) which equals to zero in our target setting. However, our approach says that these particles start, on
average, from the point $x_{start}$ which does not coincide with $\bar{x}_{tot}(0)$ in the general case. That is, our approach does not confirm
the validity of the Wigner-time concept in the general case. With taking into account of (\ref{3030}), the relationship between the asymptotic
transmission group time $\tau_{as}(k)$ introduced in our approach and the Wigner phase time $\tau_{ph}$ introduced in \cite{Ol3} on the basis of
the standard model of the process can be written as follows,
\[\tau_{as}(k)=\tau_{ph}(k)-t_{dep}(k).\]

For $L=0$, when the two-barrier system is reduced to a single rectangular barrier of width $D$, we have (see \cite{Ch1})
\[\tau_{as}(k)=\frac{4m}{\hbar k\kappa}\ooa\frac{\left[k^2+\kappa_0^2\sinh^2\left(\kappa D/2\right)\right]
\left[\kappa_0^2\sinh(\kappa D)-k^2 \kappa D\right]} {4k^2\kappa^2+ \kappa_0^4\sinh^2(\kappa D)};
\]
\begin{eqnarray} \label{304}
x_{start}(k)= -2\frac{\kappa_0^2}{\kappa}\ooa \frac{(\kappa^2-k^2)\sinh(\kappa D)+k^2 \kappa D \cosh(\kappa D)} {4k^2\kappa^2+
\kappa_0^4\sinh^2(\kappa D)}.
\end{eqnarray}
where $\kappa_0=\sqrt{2mV_0}/\hbar$ (note, focusing on the Hartman effect we assumed that $V_0>0$; however, the formalism presented is valid also
for $V_0<0$ when both $\kappa_0$ and $\kappa$ are purely imagine quantities).

As is seen from (\ref{304}), $x_{start}\to 0$ in the opaque-barrier limit, i.e., when $\kappa D\to\infty$ (providing that the requirements
(\ref{3029}) are fulfilled). Thus, in this limit, the above-mentioned assumption that underlies the concept of the Wigner time is well justified
-- $\tau_{ph}(k)\approx\tau_{as}(k)\approx \frac{2m}{\hbar k\kappa(k)}$. The fact that the Wigner time $\tau_{ph}(k)$ does not depend, in this
limit, on the width of the rectangular barrier is known as the Hartman effect.

Note that in the limit of opaque barriers the equality $\tau_{ph}(k)\approx\tau_{as}(k)\approx \frac{2m}{\hbar k\kappa(k)}$ holds also for the
two-barrier system when $L\neq 0$; that is, again, in this limit $x_{start}\to 0$. The main peculiarity of the two-barrier system is that now
$\tau_{ph}(k)$ does not depend not only on the width of two identical barriers, but also on the distance $L$ between them -- the generalized
Hartman effect \cite{Ol3}. So that the existence of both Hartman effects predicted on the basis of the standard model of the process is also
confirmed by our concept of the asymptotic group times derived for the transmission subprocess.

However, unlike the standard model ours does not associate these effects with superluminal velocities of a particle in the region $[a_1,b_2]$ (see
Section \ref{Hart}). In this region, the velocity of a tunneling particle with a definite energy is associated, according to our approach, with
the dwell time and local group time, rather than with the asymptotic group time.

\section{Dwell times for transmission and reflection} \label{loc}

Our next step is to consider the stationary scattering problem and introduce the dwell times for both subprocesses. For the two-barrier system the
dwell times $\tau^{dwell}_{tr}$ and $\tau^{dwell}_{ref}$ for transmission and reflection, respectively, are defined as follows
\begin{eqnarray} \label{305}
\tau^{dwell}_{tr}= \frac{m}{\hbar k T_{two}}\int_{a_1}^{b_2}\left|\psi_{tr}(x,k)\right|^2dx\equiv
\tau^{(1)}_{tr}+\tau^{gap}_{tr}+\tau^{(2)}_{tr},\\
\tau^{dwell}_{ref}=\frac{m}{\hbar k R_{two}}\int_{a_1}^{x_c} \left|\psi_{ref}(x,k)\right|^2dx\equiv \tau^{(1)}_{ref}+\tau^{gap}_{ref};\nonumber
\end{eqnarray}
here $\tau^{(1)}_{tr}$ and $\tau^{(1)}_{ref}$ describe the left rectangular barrier located in the interval $[a_1,b_1]$; $\tau^{gap}_{tr}$ and
$\tau^{gap}_{ref}$ characterize the free space $[b_1,a_2]$; $\tau^{(2)}_{tr}$ relates to the right rectangular barrier located in the interval
$[a_2,b_2]$.

Calculations yield (see Section \ref{back})
\begin{eqnarray}\label{400}
\tau^{(1)}_{tr}=\tau^{(2)}_{tr}=\frac{m}{4\hbar k\kappa^3}\left[2\kappa
d(\kappa^2-k^2)+\kappa_0^2\sinh(2\kappa d)\right],\\
\tau^{gap}_{tr}=\frac{m}{\hbar k^2T}\left[k
L(1+R)+4\eta\sqrt{R}\sin\left(\frac{kL}{2}\right)\sin\left(J+\frac{kL}{2}\right)\right],\nonumber\\
\tau^{(1)}_{ref}=\frac{mT_{two}|P|^2}{2\hbar k\kappa^3}\Big\{2\kappa
d\left[\kappa^2-k^2-\kappa_0^2\cos(kL)\right]+4k\kappa\sin(kL)\sinh^2(\kappa d)\nonumber\\
+\left[\kappa_0^2-(\kappa^2-k^2)\cos(kL)\right]\sinh(2\kappa d) \Big\},\ooo \tau^{gap}_{ref}(k)=\frac{mT_{two}|P|^2}{\hbar
k^2}\Big[kL-\sin(kL)\Big];\nonumber
\end{eqnarray}
here $|P|^2=[1+R-2\eta\sqrt{R}\sin(J+kL)]/T$ (see Exp. (\ref{6})).

Note that $\tau^{dwell}_{tr}(k)\neq\tau^{dwell}_{ref}(k)$ while $\tau^{as}_{tr}(k)=\tau^{as}_{ref}(k)$. Another feature is that
$\tau^{(2)}_{tr}=\tau^{(1)}_{tr}\equiv\tau^{bar}_{tr}$ (see (\ref{108})). If $\tau^{left}_{tr}$ and $\tau^{right}_{tr}$ denote the transmission
dwell times for the intervals $[a_1,x_c]$ and $[x_c,b_2]$, respectively, then
\begin{eqnarray} \label{401}
\tau^{left}_{tr}=\tau^{right}_{tr}=\tau^{bar}_{tr}+\tau^{gap}_{tr}/2= \tau^{dwell}_{tr}/2.
\end{eqnarray}
That is, the (stationary) transmission time obeys the natural physical requirement: for any barrier possessing the mirror symmetry, this
characteristic time must be the same for both symmetrical parts of the barrier.

For comparison we also present for this two-barrier system the additive characteristic time $\tau_{dwell}$ defined in the spirit of Buttiker's
dwell time \cite{But}:
\begin{eqnarray} \label{306}
\tau_{dwell}=\frac{m}{\hbar k}\int_{a_1}^{b_2}\left|\Psi_{tot}(x,k)\right|^2dx\equiv \tau_{tot}^{(1)}+\tau_{tot}^{gap}+\tau_{tot}^{(2)};
\end{eqnarray}
here the contributions $\tau_{tot}^{(1)}$, $\tau_{tot}^{(2)}$ and $\tau_{tot}^{gap}$ describe, respectively, the left and right barriers as well
as the gap between them. Calculations yield
\begin{eqnarray} \label{402}
\tau_{tot}^{(1)}=\frac{mT_{two}}{4\hbar k\kappa^3T}\bigg\{2\kappa
d\left[(\kappa^2-k^2)(1+R)+2\sqrt{R}\kappa_0^2\sin(J+kL)\right]+\\
\left[\kappa_0^2(1+R)+2\sqrt{R}(\kappa^2-k^2)\sin(J+kL)\right]\sinh(2\kappa
d)\nonumber\\
+8k\kappa\sqrt{R}\cos(J+kL)\sinh^2(\kappa d)\bigg\}\nonumber\\
\tau_{tot}^{gap}=\frac{mT_{two}}{\hbar k^2 T}\left[kL(1+R)+2\eta\sqrt{R}\sin(J+kL)\sin(kL)\right],\ooo
\tau_{tot}^{(2)}=\tau_{tr}^{(2)}T_{two}.\nonumber
\end{eqnarray}
As is seen, unlike $\tau^{dwell}_{tr}$ the Buttiker dwell time $\tau_{dwell}$ does not possess the property (\ref{401}). Besides, from Exps.
(\ref{400}) and (\ref{402}) it follows that $\tau_{dwell}$ describes neither transmitted nor reflected particles.

Let us consider the limit of opaque barriers when $\kappa d\to\infty$ and $cos^2(J_{(\infty)}+kL)\gg T^2/4R$; here
$J_{(\infty)}=\arctan(\theta_{(-)})$. In this case, we will imply that $d\to\infty$ but other parameters are fixed. Omitting the exponentially
small terms in the expressions for each scattering time, for all three contributions in Exp. (\ref{400}) to $\tau^{dwell}_{tr}$ as well as for
$\tau^{dwell}_{ref}$ we obtain
\begin{eqnarray*}
\tau^{(1)}_{tr}=\tau^{(2)}_{tr}\approx\frac{m\kappa_0^2}{4\hbar k\kappa^3}e^{2\kappa d},\ooo \tau^{gap}_{tr}\approx \frac{m\theta_{(+)}^2}{2\hbar
k^2}\left[kL+ 2\sin\left(\frac{kL}{2}\right)\sin\left(J_{(\infty)}+\frac{kL}{2}\right)\right]e^{2\kappa d};\\
\tau^{dwell}_{ref}\approx\tau^{(1)}_{ref}\approx \frac{m}{2\hbar
k\kappa^3\theta_{(+)}^2}\left[\kappa_0^2-(\kappa^2-k^2)\cos(kL)+k\kappa\sin(kL)\right].
\end{eqnarray*}
As regards the Buttiker's dwell time $\tau_{dwell}$, in this limit $\tau_{dwell}$ is determined by $\tau_{tot}^{(1)}$ which saturates in this
limit.

That is, the dwell time $\tau^{dwell}_{ref}$ for reflection and $\tau_{as}$ (as well as $\tau_{dwell}$ and $\tau_{ph}$ that appear in the standard
approach) saturate in the opaque-barrier limit, while $\tau^{dwell}_{tr}$ grows exponentially in this case! Thus, two scattering times,
$\tau_{as}$ and $\tau^{dwell}_{tr}$, that describe in the nonlinear model of the process the transmission subprocess, demonstrate a qualitatively
different behaviour in the opaque-barrier limit. Unlike the asymptotic group time $\tau_{as}$, the dwell time $\tau^{dwell}_{tr}$ does not lead to
the Hartman effects. In particular, for the two-barrier system it depends on $L$ in the opaque-barrier limit (see the expression for
$\tau^{gap}_{tr}$).

\section{Numerical results for characteristic times} \label{Hart}

So, we have introduced three characteristic times for the transmission subprocess: in the stationary case, this is the dwell time
$\tau^{dwell}_{tr}$ that characterizes its dynamics in the interval $[a_1,b_2]$ occupied by the two-barrier system; for wave packets, they are the
{\it local} group time $\tau_{tr}^{loc}$ which, too, characterizes its dynamics in the interval $[a_1,b_2]$, as well as the {\it asymptotic} group
time $\tau^{as}_{tr}$ that characterizes its dynamics in the asymptotically large interval $[0,b_2+\Delta X]$. Our next step is to compare these
characteristic times for the rectangular barrier of width $D=2d$ (that is, one has to take $L=0$ in the corresponding expressions for the
two-barrier system) with the Buttiker dwell time $\tau_{dwell}$ and Wigner phase time $\tau_{ph}$.

\begin{figure}[t]
\begin{center}
\includegraphics[width=8.0cm,angle=0]{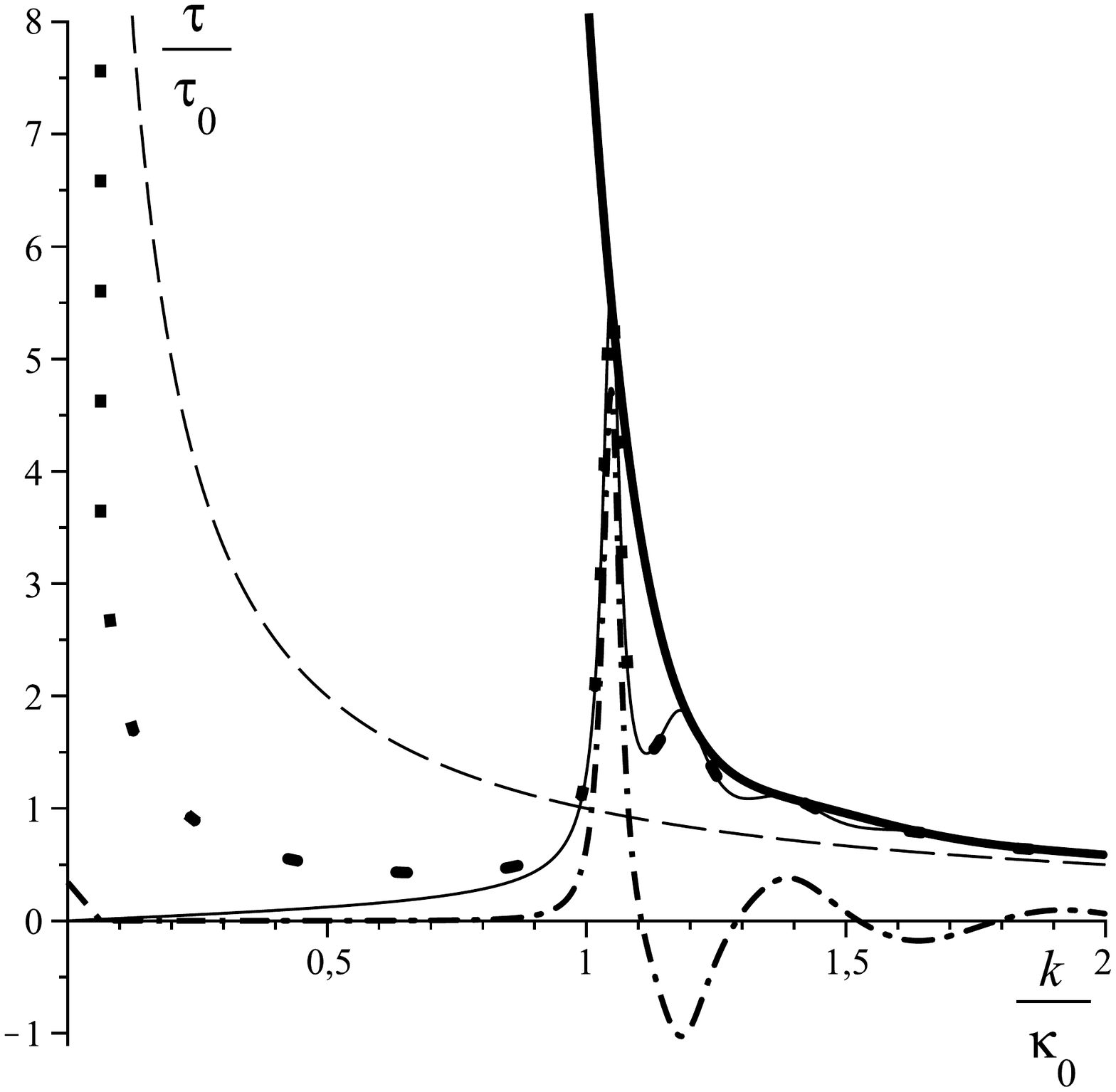}
\end{center}
\caption{$\tau^{dwell}_{tr}/\tau_0$ (bold full line), $\tau_{dwell}/\tau_0$ (full line), $\tau_{ph}/\tau_0$ (dots), $\tau_{dep}/\tau_0$ (dash-dot)
and $\tau_{free}/\tau_0$ (broken line) as functions of $k$ for a system with $L=0$ and $2\kappa_0 d =3\pi$ (see also fig.~3 in \cite{But}).}
\label{fig.1}
\end{figure}

\begin{figure}[t]
\begin{center}
\includegraphics[width=8.0cm,angle=0]{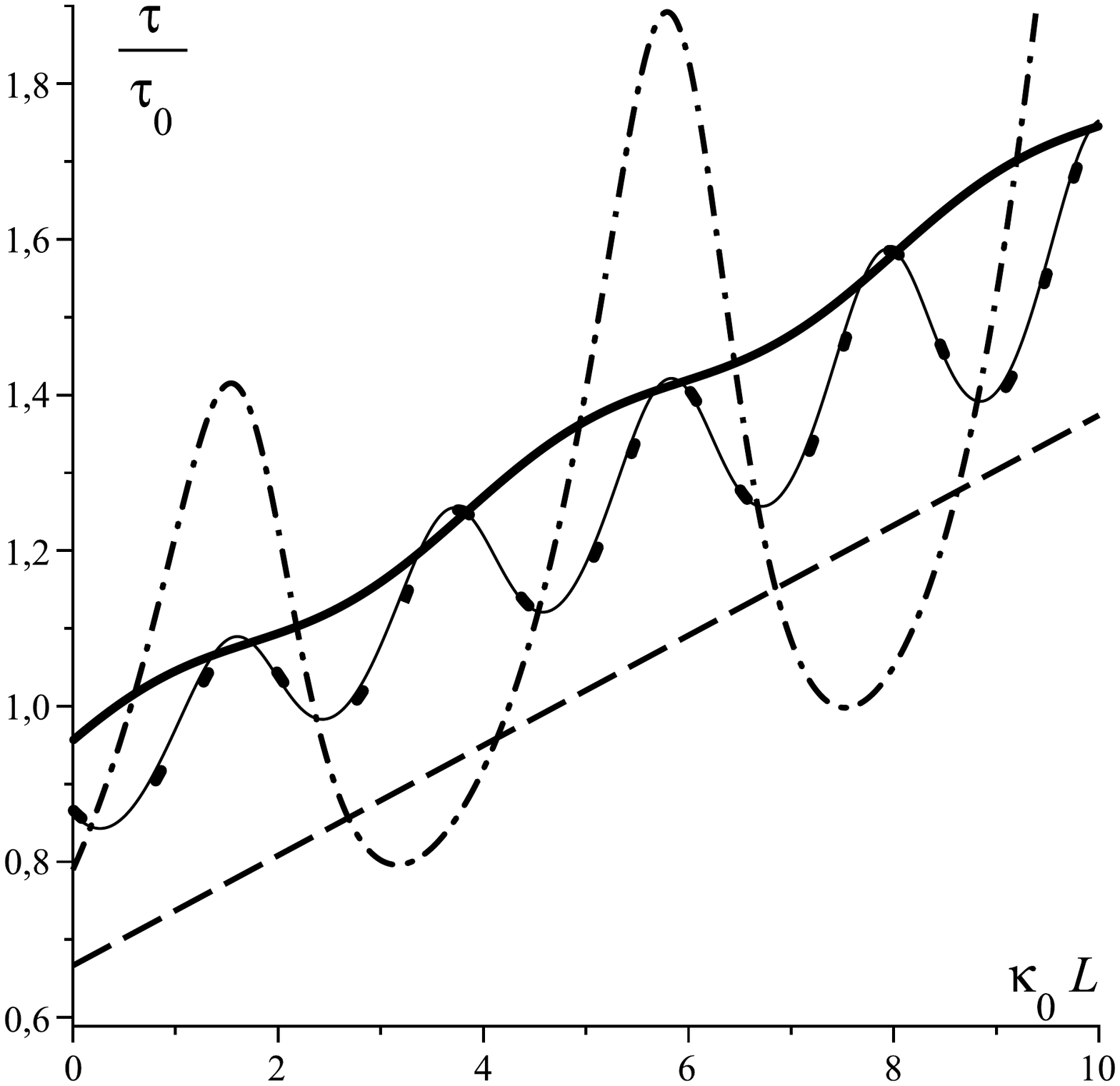}
\end{center}
\caption{$\tau^{dwell}_{tr}/\tau_0$ (bold full line), $\tau_{dwell}/\tau_0$ (full line), $\tau_{ph}/\tau_0$ (dots), $\tau_{as}/\tau_0$ (dash-dot)
and $\tau_{free}/\tau_0$ (broken line) as functions of $L$ for $2\kappa_0 d =3\pi $ and $k=1.5 \kappa_0$.} \label{fig.2}
\end{figure}

\begin{figure}[t]
\begin{center}
\includegraphics[width=8.0cm,angle=0]{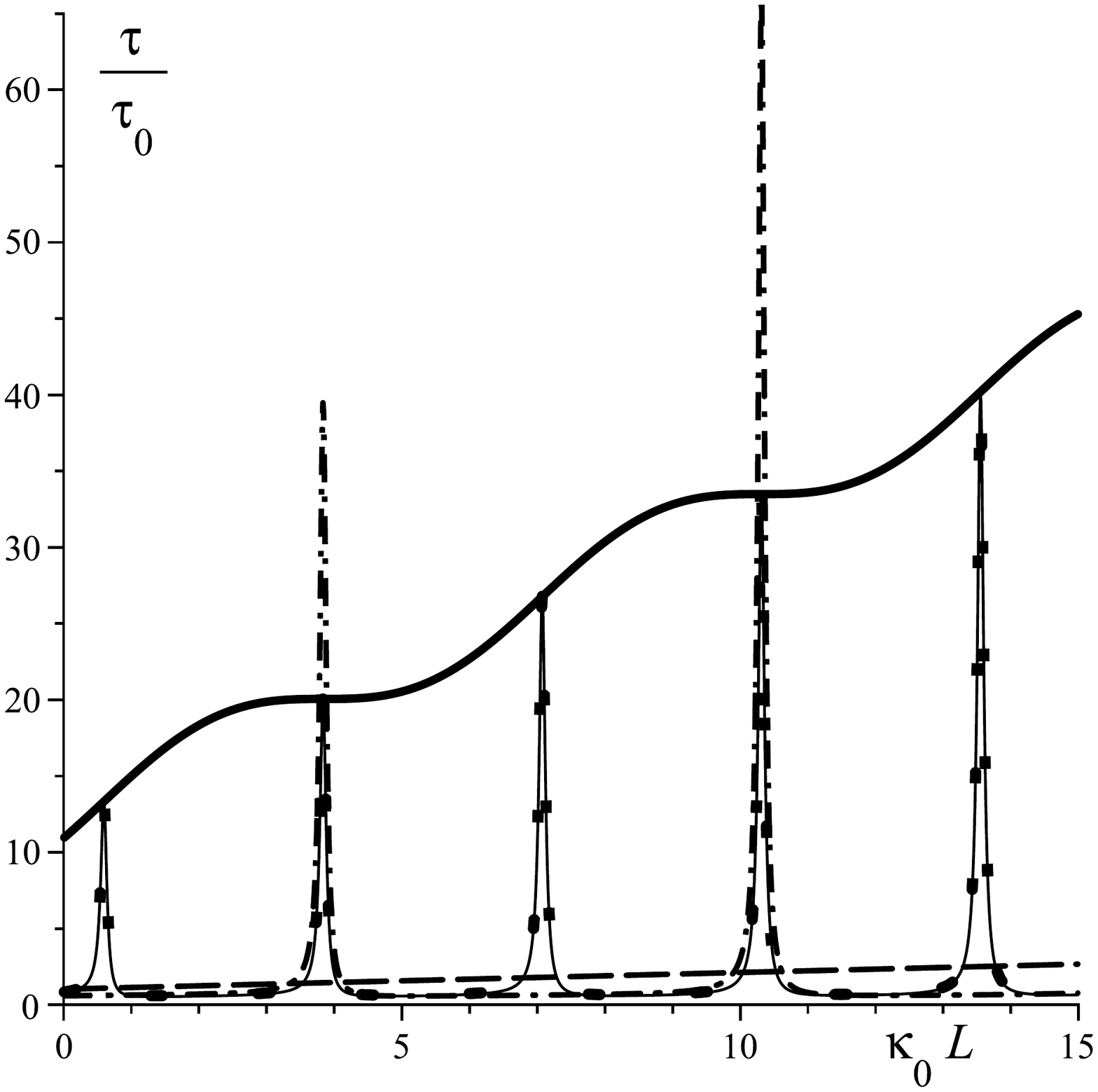}
\end{center}
\caption{$\tau^{dwell}_{tr}/\tau_0$ (bold full line), $\tau_{dwell}/\tau_0$ (full line), $\tau_{ph}/\tau_0$ (dots), $\tau_{as}/\tau_0$ (dash-dot)
and $\tau_{free}/\tau_0$ (broken line) as functions of $L$ for $2\kappa_0 d =3\pi $ and $k=0.97 \kappa_0$.} \label{fig.3}
\end{figure}

As is known, in the standard approach $\tau_{ph}$ diverges and $\tau_{dwell}$ diminishes in the low energy region, but both these quantities
coincide with each other, in the high energy region (see, e.g., fig.~3 in \cite{But}). Figs.~\ref{fig.1}-\ref{fig.3} show that in our model the
same behavior is manifested by the asymptotic group time $\tau_{as}$ (in all these figures, the quantity $\tau_{free}/\tau_0$, where
$\tau_{free}=m D/\hbar k$ and $\tau_0=mD/\hbar\kappa_0$, is presented as a 'reference' one). Unlike the standard time scales $\tau_{ph}$ and
$\tau_{dwell}$, as well as $\tau_{as}$, the transmission dwell time $\tau^{dwell}_{tr}$ never leads to 'anomalously' short tunneling times.
As is seen from fig.~\ref{fig.1}, all the analyzed time scales (of course, excluding $\tau_{dep}$) approach the free-passage time $\tau_{free}$ in
the high energy region. However, in the low energy region $$\tau^{dwell}_{tr}\gg \tanh^2\left(\frac{\kappa_0D}{2}\right)\cdot\tau_{ph}\approx
\tau_{as} \approx\frac{2}{\kappa_0D}\tanh\left(\frac{\kappa_0D}{2}\right)\cdot\tau_{free} \gg \tau^{dwell}_{ref}\approx\tau_{dwell}.$$

Note that, on the $k$-axis the quantity $\tau_{dep}$ changes its sign at the points located between resonant points, where $R_{two}=0$. At
resonance points, the function $|\tau_{dep}(k)|$, like the dwell time $\tau_{dwell}$ and the Wigner phase time $\tau_{ph}$, takes extreme values
(see figs.~\ref{fig.1}--\ref{fig.3}). At these points, the Buttiker dwell time $\tau_{dwell}(k)$ coincides with $\tau^{dwell}_{tr}(k)$ and takes
maximal values in the vicinity of these points. Note that the behavior of $\tau_{as}(k)$ is more complicated: like $\tau_{dwell}(k)$ it takes
maximal values at the even resonance points (in this case $\tau_{as}(k)\approx 2\tau_{dwell}(k)$, while $\tau_{as}(k)$ has no maxima at the odd
resonance points, including the first (lowest) resonance point (see fig.~\ref{fig.3})).

The CMs of the wave packet $\psi_{tr}(x,t)$, peaked on the $k$-scale at the resonance points with the even numbers, starts earlier
($\tau_{dep}(k)<0$) than the CM of the total wave packet $\Psi_{tot}(x,t)$. While at the resonance points with odd numbers we find the opposite
situation. Moreover, at such energy points, the local maxima of the function $\tau_{ph}(k)$ transform into the local minima of the function
$\tau_{as}(k)=\tau_{ph}(k)-t_{tr}^{dep}(k)$.

When $L\neq 0$ and $E>V_0$, all scattering times show the tendency to increase in the limit $L\to\infty$ (see fig.~\ref{fig.2}). However, in the
tunneling regime ($E<V_0$), only the dwell time $\tau^{dwell}_{tr}(k)$ monotonously increases (see fig.~\ref{fig.3}). Other characteristic times
do not in fact depend on $L$ in the opaque-barrier limit (the generalized Hartman effect). Moreover, for $\tau_{as}(k)$ this takes place also at
the odd resonance points).

So, for the two-barrier system with opaque rectangular barriers, the dwell time $\tau^{dwell}_{tr}$ is much larger than the asymptotic group time
$\tau_{as}$ which does not depend on $L$ in this case. However, this fact does not at all mean that our approach leads to mutually contradictory
tunneling time concepts, with one of them violating special relativity. We have to remember that, unlike $\tau^{dwell}_{tr}$, the asymptotic
characteristic time $\tau_{as}$ is not a tunneling time. In order to demonstrate the difference between these two characteristic times let us
consider in the time-dependent case the function ${\bar{x}}_{tr}(t)$ to describe scattering the Gaussian wave packet (\ref{200}) on the
rectangular potential barrier (i.e., now $L=0$): $l_0=10nm$, $\bar{E}=(\hbar\bar{k})^2/2m=0.05eV$, $a_1=200nm$, $b_2=215nm$, $V_0=0.2eV$.

\begin{figure}[h]
\begin{center}
\includegraphics[width=8.0cm,angle=-90]{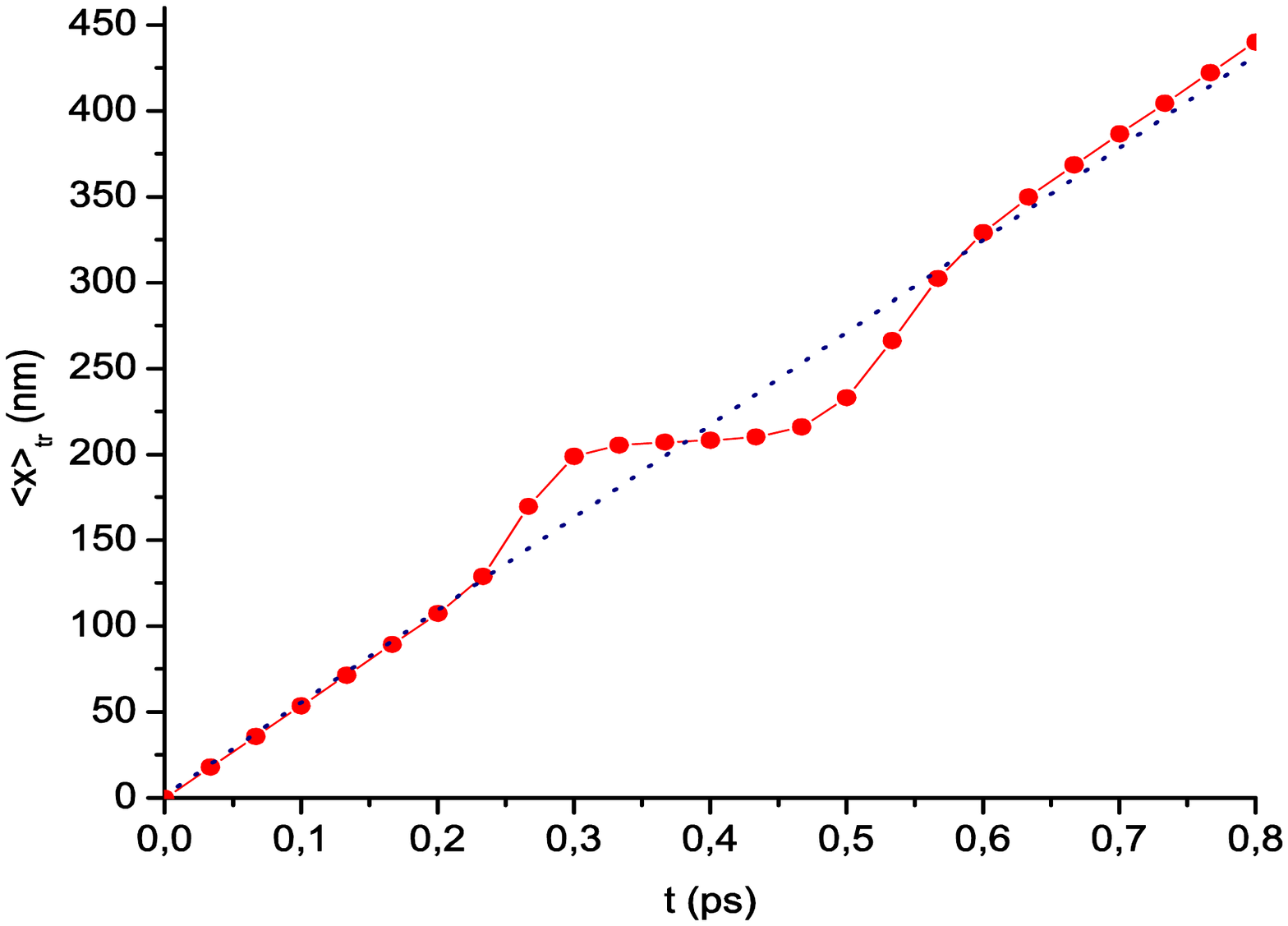}
\end{center}
\caption{The CM's positions for $\psi_{tr}(x,t)$ (circles) and for the corresponding freely moving reference wave packet (dashed line) as
functions of time $t$.} \label{fig:fig5a1}
\end{figure}

For this case calculations yield $\tau^{loc}_{tr}\approx 0,155ps$, $\tau^{as}_{tr}\approx 0,01ps$, $\tau_{free}\approx 0,025ps$ (see
fig.~\ref{fig:fig5a1}). This figure shows explicitly a qualitative difference between the local $\tau_{tr}^{loc}$ and asymptotic $\tau_{tr}^{as}$
transmission group times. While the former gives the time spent by the CM of this packet in the region $[a_1,b_2]$, the latter describes the
influence of the barrier on this CM in the asymptotically large interval $[0,b_2+\Delta X]$. Thus, the quantity $\tau^{as}_{tr}-\tau_{free}$ is a
time delay which is acquired by this CM in the course of the whole scattering process; $\tau_{free}=mD/\hbar k$. It describes the relative motion
of the CMs of the {\it transmitted} wave packet and the CM of a reference packet that moves freely at all stages of scattering and departs from
the point $x=0$ at the time $\tau_{dep}$. When the barrier is opaque, $\tau_{dep}$ coincides approximately with the departure time of the total
wave packet $\Psi_{tot}(x,t)$.

Thus, the influence of an opaque rectangular barrier on the transmitted wave packet has a complicated character. The {\it local} group time
$\tau_{tr}^{loc}$, together with the dwell time $\tau^{dwell}_{tr}$, says that the opaque barrier retards the motion of the CM in the barrier
region, while the {\it asymptotic} group time $\tau_{tr}^{as}$ tells us that, in the course of the whole scattering process, the total influence
of the opaque barrier on the transmitted wave packet has an accelerating character: at the final stage of scattering, this packet moves ahead the
schedule of the reference packet; $\tau^{as}_{tr}-\tau_{free}\approx -0,015ps$.

Note, for any finite value of $l_0$, the velocity of the {\it CM} of the wave packet $\psi_{tr}(x,t)$ is constant only at the initial and final
stages of scattering. However, when the value of $l_0$ is large enough (a quasi-monochromatic wave packet) this takes place also at the very stage
of scattering, when the CM of this packet moves inside the region $[a_1,b_2]$ while its leading and trailing fronts are far beyond this region. At
this stage, only the main harmonic $\bar{k}$ determines the input and output probability flows at the point $x_c$. As a result, these flows
balance each other, and hence the norm ${\textbf{T}}$ is constant at this stage. In this case the {\it local} group time $\tau^{as}_{tr}$
approaches the dwell time $\tau^{dwell}_{tr}$; in the opaque-barrier limit, both predict the effect of retardation of tunneling particles in the
region $[a_1,b_2]$.

Another situation arises in this limit when either the leading or trailing front of the wave packet $\psi_{tr}(x,t)$ crosses the point $x_c$. In
the first case, due to the interference effect discussed in Section \ref{alt}, this point serves as a 'source' of particles, what leads to the
acceleration of its CM located at this stage to the left of the barrier. While in the second case this spatial point effectively acts as an
'absorber' of to-be-transmitted particles; what leads again to the acceleration of this CM which is located now to the right of the barrier (see
fig~\ref{fig:fig5a1}). In the case presented on this figure, the velocity of the CM, prior to its entering into the barrier region and after its
exit from this region, is larger three times compared with its velocity at the first and final stages of scattering. In the last analysis, this
effect leads to the saturation of the asymptotic transmission group time, in the opaque-barrier limit.

Thus, in our approach the usual Hartman effect does not at all mean that a particle tunnels through the opaque potential barrier with a
superluminal velocity. Rather, it means that the subensemble of tunneling particles is accelerated by the opaque barrier when the average distance
between particles and the nearest boundary of the barrier equals to $l_0/2$. And what is important is that this acceleration does not lead to
superluminal velocities of a particle.

One has to bear in mind that the accuracy of determining the coordinates of the leading and trailing fronts of the wave packet $\psi_{tr}(x,t)$ is
proportional to its width. Thus, the wider is the packet, the larger is the size $L_{accel}$ of spatial intervals where the velocity of its CM
grows. This means, in particular, that $L_{accel}$ grows in the opaque-barrier limit $d\to\infty$. This is so because the propagation of a
quasi-monochromatic wave packet in the asymptotically large interval $[0,b_2+\Delta X]$, in the opaque-barrier limit $d\to\infty$, implies the
validity of the inequalities $a_1, \Delta X\gg l_0\gg D$ (\ref{3029}). Thus, in this limit, the wave-packet's width $l_0$ (and hence $L_{accel}$)
grows together with $D=2d+L$. That is, the opaque-barrier limit leads to the growth of the spatial and temporal scales of the curve
$\bar{x}_{tr}(t)$, rather than to the unbounded growth of the average velocity of particles passing trough the opaque barrier.

\section{Conclusion}\label{conclude}

We showed that the superposition principle is violated in the quantum mechanical process of scattering a particle on a one-dimensional potential
barrier as well as in the process of scattering the electromagnetic wave on a quasi-one-dimensional layered structure. The barrier and the
structure, dividing the incident wave into two parts (transmitted and reflected), play the role of nonlinear elements in the corresponding
scattering problems. Thus, both in QM and in CED, these scattering phenomena are nonlinear, in reality. As a consequence, the standard (linear)
models of these two scattering processes are not adequate to them, which makes the study of the temporal aspects of each of these two processes an
intractable problem. This explains why the TTP remains a controversial issue right up to our days.

By the example of scattering a quantum particle on a one-dimensional potential barrier we present a new, nonlinear model of this process. In
particular, we find the (stationary) wave functions for the transmission and reflection subprocesses which allow one to study them at all stages
of scattering. On their basis we define the dwell times and ("phase") group times for each subprocess. As was shown, only the dwell time and the
local group time, defined for the transmission subprocess, can be treated as the transmission (tunneling) time, that is, the time spent by a
transmitted (tunneled) particle in the barrier region. These times correlate with each other and do not lead to the Hartman paradox. As it follows
from our model, a direct measurement of the tunneling time is impossible.

From our point of view the Hartman paradox stands alongside with the Schr\"odinger cat paradox and the mystery of the double-slit experiment. They
have a common root cause. According to our approach, the Schr\"odinger cat paradox is not a measurement problem. It must be resolved at the
micro-level: in quantum mechanics, the decay model of radioactive nuclei should be nonlinear, like the model of scattering of a quantum particle
on a one-dimensional potential barrier.


\begin{thebibliography}{99}
\bibitem{Soko}
D. Sokolovski: The Salecker-Wigner-Peres clock, Feynman paths, and a tunnelling time that should not exist.
\end{thebibliography}
\end{document}